\input harvmac
\input epsf

\def\d{\Lambda}

\def\etal{ et al.}



\Title{}{\vbox{\centerline{Supernova Constraints on a holographic
dark energy model}}}

\centerline{Qing-Guo Huang$^{1,2}$ and Yungui Gong$^{3}$ }
\medskip
\centerline{$^{1}$\it Institute of Theoretical Physics}
\centerline{\it Academia Sinica, P.O. Box 2735} \centerline{\it
Beijing 100080}
\medskip
\centerline{$^{2}$\it Interdisciplinary Center for Theoretical
Study,} \centerline{\it University of Science and Technology of
China,} \centerline{\it Hefei, Anhui 230026, P. R. China}
\medskip
\centerline{$^{3}$\it Center for Relativity and Astrophysics and
College of Electronic Engineering, } \centerline{\it Chongqing
University of Posts and Telecommunications, } \centerline{\it
Chongqing 400065, China}
\medskip
\centerline{\tt huangqg@itp.ac.cn} \centerline{\tt
gongyg@cqupt.edu.cn}
\bigskip

In this paper, we use the type Ia supernova data to constrain the
model of holographic dark energy. For $d=1$, the best fit result
is $\Omega^0_m = 0.25$, the equation of the state of the
holographic dark energy $w^0_\d = -0.91$ and the transition
between the decelerating expansion and the accelerating expansion
happened when the cosmological red-shift was $z_T=0.72$. If we set
$d$ as a free parameter, the best fit results are $d=0.21$,
$\Omega_m^0=0.46$, $w_\d^0=-2.67$, which sounds like a phantom
today, and the transition redshift is $z_T=0.28$.

\Date{March, 2004}

\nref\agr{A. G. Riess et al., Astron. J. 116 (1998) 1009; S.
Perlmutter et al., APJ 517 (1999) 565. }

\nref\tonbar{J. L. Tonry et al., APJ 594 (2003) 1; R. A. Knop et
al., astro-ph/0309368; B. J. Barris et al., APJ 60 (2004) 571.}

\nref\spna{A. G. Riess et al, astro-ph/0402512. }

\nref\cmb{P. de Bernardis et al., Nature 404 (2000) 955; S. Hanany
et al., APJ 545 (2000) L5; D. N Spergel et al., APJ Supp. 148
(2003) 175.}

\nref\dark{C. Wetterich, Nucl. Phys. B 302 (1988) 668; B. Ratra
and P. J. E. Peebles, Phys. Rev. D 37 (1988) 3406; R. R. Caldwell,
R. Dave and P. J. Steinhardt, Phys. Rev. Lett. 80 (1998) 1582; I.
Zlatev, L. Wang and P. J. Steinhardt, Phys. Rev. Lett. 82 (1999)
896.}

\nref\review{ V. Sahni and A. A. Starobinsky, Int. J. Mod. Phys. D
9 (2000) 373; T. Padmanabhan, Phys. Rep. 380 (2003) 235.}

\nref\thooft{ G. 't Hooft, Salamfestschrift: a collection of
talks, Eds. A. Ali, J. Ellis and S. Randjbar-Daemi, World
Scientific, 1993.}

\nref\susskind{ L. Susskind, J. Math. Phys. 36 (1994) 6377.}

\nref\ckn{A. Cohen, D. Kaplan and A. Nelson, Phys. Rev. Lett. 82
(1999) 4971.}

\nref\hsu{S. D. H. Hsu, hep-th/0403052. }

\nref\horvat{R. Horvat, astro-ph/0404204.}

\nref\ph{W. Fischler and L. Susskind, hep-th/9806039; N. Kaloper
and A. Linde, Phys. Rev. D 60 (1999) 103509; R. Bousso, Rev. Mod.
Phys. 74 (2002) 825. }

\nref\ml{Miao Li, hep-th/0403127. }

\nref\tbf{T. Banks and W. Fischler, astro-ph/0307459.}

\nref\tegmark{M. Tegmark et al., Phys. Rev. D 69 (2004) 103501.}

\nref\gong{Y. Gong, astro-ph/0401207.}

\nref\indep{R. A. Daly and S. G. Djorgovski, astro-ph/0403664; U.
Alam, V. Sahni, T.D. Saini and A.A. Starobinsky, astro-ph/0311364;
U. Alam, V. Sahni and A. A. Starobinsky, astro-ph/0403687; H. K.
Jassal, J. S. Bagla and T. Padmanabhan, astro-ph/0404378; Y. Gong,
astro-ph/0405446.}

The type Ia supernova (SN Ia) observations \refs{\agr, \tonbar,
\spna} provide evidence that the expansion of our universe at the
present time appears to be accelerating which is due to dark
energy with negative pressure. The kinematic interpretation of the
relationship between SN Ia luminosity distance and red-shift is
most consistent with two distinct epoches of expansion: a recent
accelerated expansion and a previous decelerated expansion with a
transition around $z_{\rm T}\sim 0.4$ between them \spna. This is
a generic requirement of a mixed dark matter and dark energy
universe. The cosmic background microwave (CMB) observations \cmb\
hint a spatially flat universe. However, the unusual small value
of the cosmological constant, which is extremely smaller than the
estimate from the effective local quantum field theory, is perhaps
one of the biggest puzzles and deepest mysteries in modern
physics. Alternatively, many dynamical dark energy models with
cosmological constant like behavior were proposed in the
literature \dark\ and \review. For a review, see, for example
\review\ and references therein.

't Hooft \thooft\ and Susskind \susskind\ showed that the
effective local quantum field theories greatly over-count degrees
of freedom because the entropy scales extensively for an effective
quantum field theory in a box of size $L$ with UV cutoff $\d$. In
order to solve the problem, A. Cohen et al. \ckn\ proposed a
relationship between UV and IR cut-offs corresponding to the
assumption that the effective field theory describes all states of
the system excluding those for which have already collapsed to a
black hole. If the sum of the zero-point energies of all normal
modes of the fields is $\rho_\d$, we must have $L^3 \rho_\d \leq L
M_p^2$ or $\rho_\d \leq M_p^2 L^{-2}$, this means that the maximum
entropy is in the order of $S^{3/4}_{BH}$. The magnitude of the
holographic energy proposed by Cohen et al. may be the same as
that from cosmological observations. But Hsu recently pointed out
that the equation of state is not correct for describing the
accelerating expansion of our Universe in \hsu. In other words,
the original holographic energy couldn't give an accelerating
universe. The idea was later generalized to make the gravitational
constant varying with time in \horvat.

The origin of the Bekenstein-Hawking constraint on the entropy of
a black hole is the existence of the event horizon, which serves
as a natural boundary for all processes inside a black hole.
However, there is no event horizon in a non-inflationary universe
and we should replace it with the particle horizon, which has been
discussed in \ph. But there is an event horizon in the Universe
with accelerating expansion and it is a natural choice that the
event horizon acts as the boundary of the Universe. Very recently,
Li suggested that we should use the proper future event horizon of
our Universe to cut-off the large scale and bring about an
accelerating expansion of our Universe in \ml. On the other hand,
Banks and Fischler have pointed out that the the number of the
e-floldings during inflation is bounded, which is due to the bound
on the entropy, if we take the event horizon as the boundary of
our Universe and the present acceleration of the Universe is due
to an asymptotically de Sitter universe with small cosmological
constant in \tbf.

In this paper, we use the new SN Ia data compiled by Riess \etal\
to constrain the holographic dark energy model proposed by Li.
Firstly we take a short trip on the holographic energy model
proposed in \ml. According to \ml, the energy density of the
holographic dark energy is \eqn\erd{\rho_\d = 3 d^2 M^2_p
R^{-2}_h, } here we keep $d$ as a free parameter (the author of
\ml\ favored $d=1$) and $R_h$ is the proper size of the future
event horizon, \eqn\eh{R_h(t) = a(t) \int^{\infty}_{t} {dt' \over
a(t')} = a \int^{\infty}_{a} {da' \over H' a'^2}. } For a
spatially flat, isotropic and homogeneous universe with an
ordinary matter and dark energy, the Friedmann equation is
\eqn\fde{ \Omega_\d + \Omega_m = 1, \quad \Omega_m = {\rho_m \over
\rho_{cr}} \quad \hbox{and} \quad \Omega_\d={\rho_\d \over
\rho_{cr}}, } where $\rho_m$ ($\rho_\d$) is the energy density of
matter (dark energy) and the critical density $\rho_{cr} = 3 M^2_p
H^2$. Using Friedmann equation \fde\ and $\rho_m = \rho^0_m
a^{-3}=3 M^2_p H^2_0 \Omega_m^0 a^{-3}$, where we set $a_0 =1$ and
$a = (1 + z)^{-1}$, we have $\Omega_m=1-\Omega_\d=(H_0/H)^2
\Omega_m^0 a^{-3}$,
 \eqn\ah{{1 \over a H}
= a ^{1/2} {1 \over \sqrt{\Omega^0_m} H_0} ( 1 - \Omega_\d
)^{1/2}, } and \eqn\rd{\rho_\d = \Omega_\d \rho_{cr} = {\Omega_\d
\over 1 - \Omega_\d} \Omega_m \rho_{cr} = {\Omega_\d \over 1 -
\Omega_\d} \rho_m = {\Omega_\d \over 1 - \Omega_\d} \rho^0_m
a^{-3}. } Combining equations \erd\ and \rd, we find
\eqn\rhm{R_h(t) = a^{3/2} {d \over \sqrt{\Omega^0_m} H_0} \left({1
- \Omega_\d \over \Omega_\d} \right)^{1/2}. } Substituting
equations \ah\ and \rhm\ into \eh, \eqn\inq{\int^{\infty}_{x} dx'
e^{x'/2} (1 - \Omega_\d)^{1/2} = d e^{x/2} \left({1 - \Omega_\d
\over \Omega_\d} \right)^{1/2}, } where $x = \ln a$. Taking
derivative with respect to $x$ in both sides of equation \inq, we
get \eqn\dnq{\Omega'_\d = \Omega_\d (1 - \Omega_\d) ( 1 + {2 \over
d} \sqrt{\Omega_\d} ), } where the prime denotes the derivative
with respect to $x$. We can get the analytic solution of equation
\dnq\ as, \eqn\mz{\ln \Omega_\d - {d \over 2+d} \ln (1 -
\sqrt{\Omega_\d}) + {d \over 2-d} \ln (1 +\sqrt{\Omega_\d}) - {8
\over 4-d^2} \ln(d + 2 \sqrt{\Omega_\d}) = - \ln(1 + z) + y_0, }
where $y_0$ can be determined by the value of $\Omega^0_\d$
through equation \mz.

Because of the conservation of the energy-momentum tensor, the
evolution of the energy density of dark energy is governed by
\eqn\eed{{d \over da} (a^3 \rho_\d) = -3 a^2 p_\d. } Thus we
obtain \eqn\pd{p_\d = - {1 \over 3} {d \rho_\d \over d \ln a} -
\rho_\d. } Using equations \rd\ and \dnq, after a lengthy but
straightforward calculation, we find the pressure of dark energy
can be expressed as \eqn\pd{p_\d = - {1 \over 3} (1 + {2 \over d}
\sqrt{\Omega_\d}) \rho_\d, } and the equation of state of dark
energy is \eqn\md{w_\d = {p_\d \over \rho_\d} = - {1 \over 3} (1 +
{2\over d} \sqrt{\Omega_\d}), } and  \eqn\rmd{{d w_\d \over d z} =
{1 \over 3d} \sqrt{\Omega_\d} (1 - \Omega_\d) ( 1 + {2\over d}
\sqrt{\Omega_\d} ) {1 \over 1 + z}. } Since $0 \leq \Omega_\d \leq
1$, we find the equation of state of dark energy $-(1+2/d)/3 \leq
w_\d \leq -{1\over 3}$ and the evolution of $w_\d$ is slow. If we
use $\Omega^0_\d = 0.73$ and $d=1$, we obtain $w^0_\d = - 0.90$
and $dw^0_\d/dz=0.21$. In the future, our Universe will be
dominated by dark energy with $w_\d = -(1+2/d)/3$, this result is
the same as Eq. (9) in \ml. The expansion of our Universe will be
accelerating forever.

In the past, the expansion of our Universe experienced
deceleration due to domination by radiation or matter. With the
evolution, our Universe will be dominated by the dark energy and
the expansion of our Universe starts to be accelerating. This
transition happened when \eqn\acc{{\ddot a \over a} = - {1 \over 6
M^2_p} (\rho_\d + 3 p_\d + \rho_m) = 0.}  Using equations \rd\ and
\pd, we obtain \eqn\tran{\Omega^T_\d+{2\over d}\Omega^T_\d
\sqrt{\Omega^T_\d}=1. } If $d=1$, solving this equation, we find
that the turning point is corresponding to $\Omega^T_\d=0.4320$.
For $d \rightarrow \infty$, $\Omega_\d^T \simeq 1$. If
$\Omega_\d^0$ is finite, $y_0$ in equation \mz\ will be finite.
Using equation \mz\ again, we obtain the red-shift of turning
point must be $z_T \simeq -1$. This result can be understood
easily. Since $d \rightarrow \infty$ and $\Omega_\d$ must be
finite, $w_\d \rightarrow -1/3$. But the energy density of matter
will be red-sifted faster than the holographic dark energy. So in
the far future our Universe will be dominated by the dark energy
and its expansion will be accelerating because of $w_\d < -1/3$.
On the other hand, for $d \rightarrow 0$, equation \tran\ tells us
\eqn\dtran{\Omega_\d^T \simeq (d/2)^{2/3}.} Also assuming
$\Omega_\d^0$ is finite, applying equation \mz, we have $y_0 = -2
\ln 2$. Substituting equation \dtran\ into \mz, we obtain the
red-shift of turning point is $z_T \simeq 0$. In this case, $w_\d
\rightarrow - \infty$ for finite $\Omega_\d$ and the energy
density of the holographic dark energy will increase very fast.
Therefore our Universe was dominated by this dark energy and its
expansion started to be accelerating very recently, if
$\Omega_\d^0$ is finite. We show the relation between the
red-shift of the turning point and the value of the parameter $d$
for some finite $\Omega_\d^0$ in Fig. 1. This figure is consistent
with our previous analysis.
\bigskip
{\vbox{{\epsfxsize=8cm
        \nobreak
    \centerline{\epsfbox{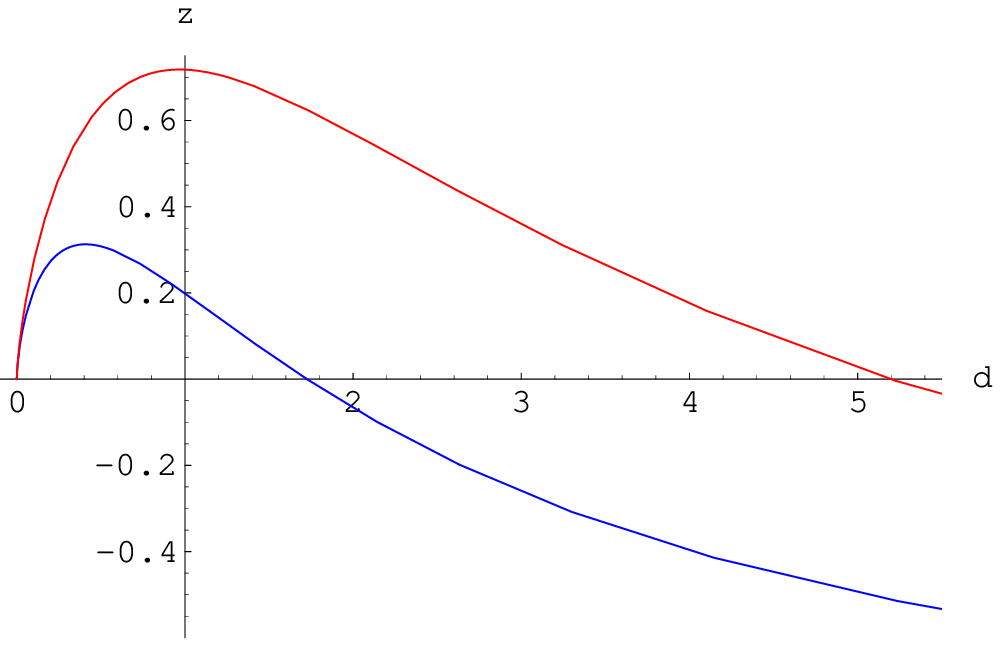}}
        \nobreak\bigskip
    {\raggedright\it \vbox{
{\bf Figure 1.} {\it $z = z_T$ is the cosmological red-shift
corresponding to the turning point between decelerating expansion
and accelerating expansion, here the blue line corresponds to
$\Omega^0_\d = 0.75$ and the red one corresponds to
$\Omega_\d^0=0.54$.}
 }}}}
    \bigskip}

The luminosity distance $d_{\rm L}$ expected in a spatially flat
Friedmann - Robertson - Walker (FRW) cosmology with mass density
$\Omega_m$ and the holographic dark energy density $\Omega_\d$ is
\eqn\lumin{\eqalign{ d_{\rm L}(z)=&c(1+z)\int^{t_0}_t {d t'\over
a(t')}=c(1+z)[(1+z)R_h(t)-R_h(t_0)] \cr
=&c\,d\,H^{-1}_0\left[\sqrt{{1-\Omega_\Lambda\over \Omega_\Lambda
\Omega_m^0}}(1+z)^{1/2}-(1-\Omega^0_m)^{-1/2}(1+z)\right] .}} In
the above derivation, we used Eqs. \eh and \rhm. The parameter
$\Omega^0_m$ of the model and the nuisance parameter $H_0$ are
determined by minimizing \eqn\lrmin{ \chi^2=\sum_i{[\mu_{\rm
obs}(z_i)-\mu(z_i)]^2\over \sigma^2_i},} where the
extinction-corrected distance moduli $\mu(z)=5\log_{10}(d_{\rm
L}(z)/{\rm Mpc})+25$ and $\sigma_i$ is the total uncertainty in
the observation. The nuisance parameter $H_0$ is marginalized. If
$d=1$, the best fit to the 157 gold SN sample in \spna\ is
$\Omega_m^0=0.25^{+0.04}_{-0.03}$ with $\chi^2=176.7$ or
$\chi^2/{\rm dof}=1.133$, and the best fit to the whole 186 gold
and silver SN sample is $\Omega_m^0=0.25\pm{0.03}$ with
$\chi^2=232.8$ or $\chi^2/{\rm dof}=1.258$. By using the best fit
$\Omega_m^0$, we find that $w^0_\d=-0.91 \pm 0.01$ and the
red-shift corresponding to transition is
$z_T=0.72^{+0.11}_{-0.13}$. For comparison, the best fit to the
gold SN sample for the $\d$-model is $\Omega_m^0=0.31\pm 0.04$
with $\chi^2=177.1 $ or $\chi^2/{\rm dof}=1.135$. So the
transition redshift for $\d$-model is $z_T=0.65^{+0.11}_{-0.10}$.
If we set $d$ as a free parameter, we find the best fit to the
gold SN sample is $\Omega_m^0=0.46^{+0.08}_{-0.13}$ and
$d=0.21^{+0.45}_{-0.14}$ with $\chi^2=173.45$ or $\chi^2/{\rm
dof}=1.119$. In this case the red-shift corresponding to
transition is $z_T=0.28^{+0.23}_{-0.13}$. The best fit contour for
$\Omega_m^0$ and $d$ is plotted in Fig. 2. The best fit to the
gold and silver SN sample is $\Omega_m^0=0.46^{+0.14}_{-0.11}$ and
$d=0.20^{+0.28}_{-0.10}$ with $\chi^2=226.4$ or $\chi^2/{\rm
dof}=1.230$. And the red-shift corresponding to transition is
$z_T=0.27^{+0.19}_{-0.14}$.

\bigskip
{\vbox{{\epsfxsize=8cm
        \nobreak
    \centerline{\epsfbox{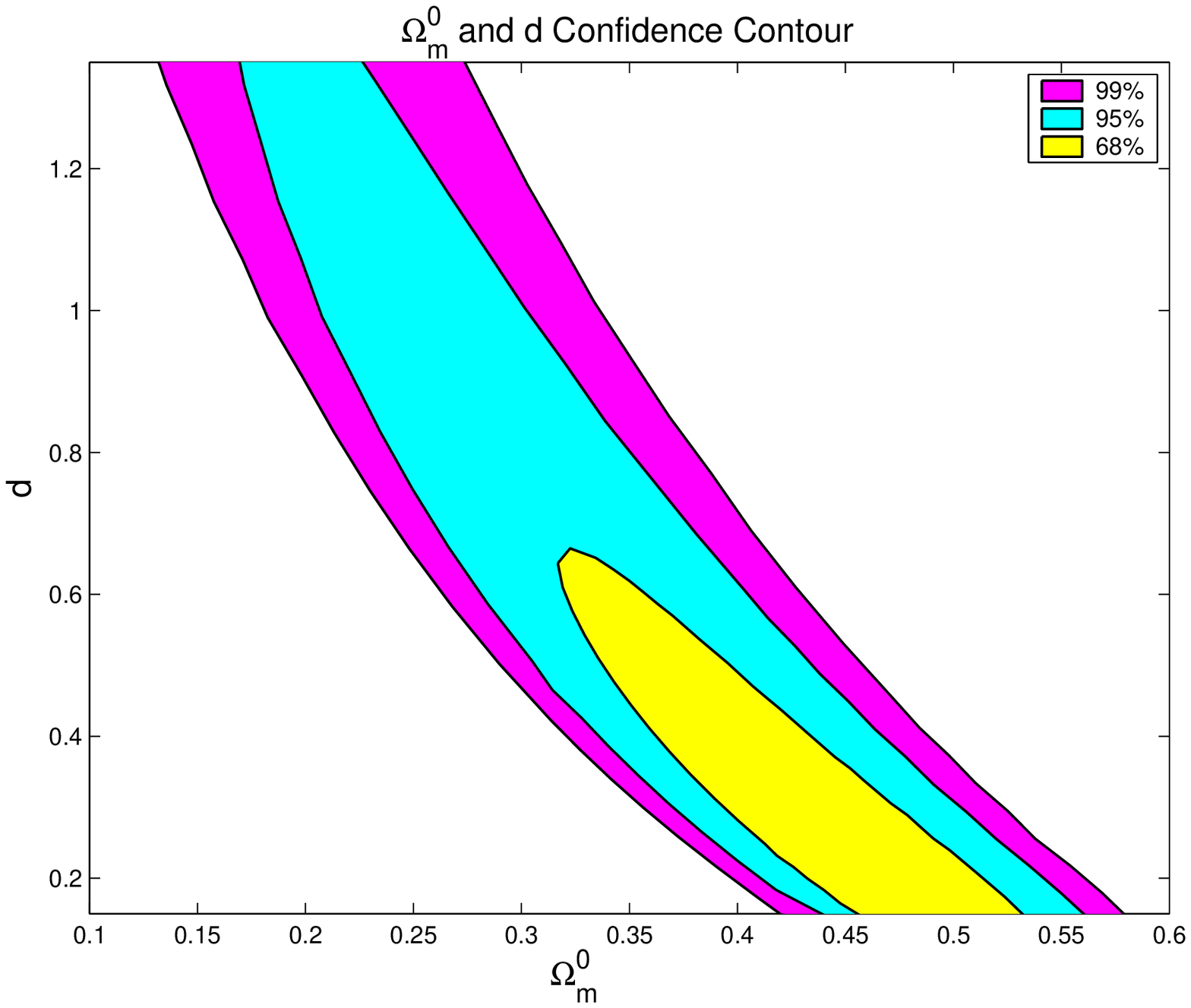}}
        \nobreak\bigskip
    {\raggedright\it \vbox{
{\bf Figure 2.} {\it The best fit contour for $\Omega_m^0$ and $d$
to the gold sample SNe.}
 }}}}
    \bigskip}

Using the fit parameters $\Omega^0_\d = 0.75$ and $d=1$, we get
$y_0 = -1.67$. With the best fit parameters $\Omega^0_\d = 0.54$
and $d=0.21$, we get $y_0 = -1.47$. Combining equations \mz, \md\
and \lumin, we show the equation of state of the dark energy
$w_\d$ in Fig. 3 and the extinction-corrected distance moduli in
Fig. 4.
\bigskip
{\vbox{{\epsfxsize=8cm
        \nobreak
    \centerline{\epsfbox{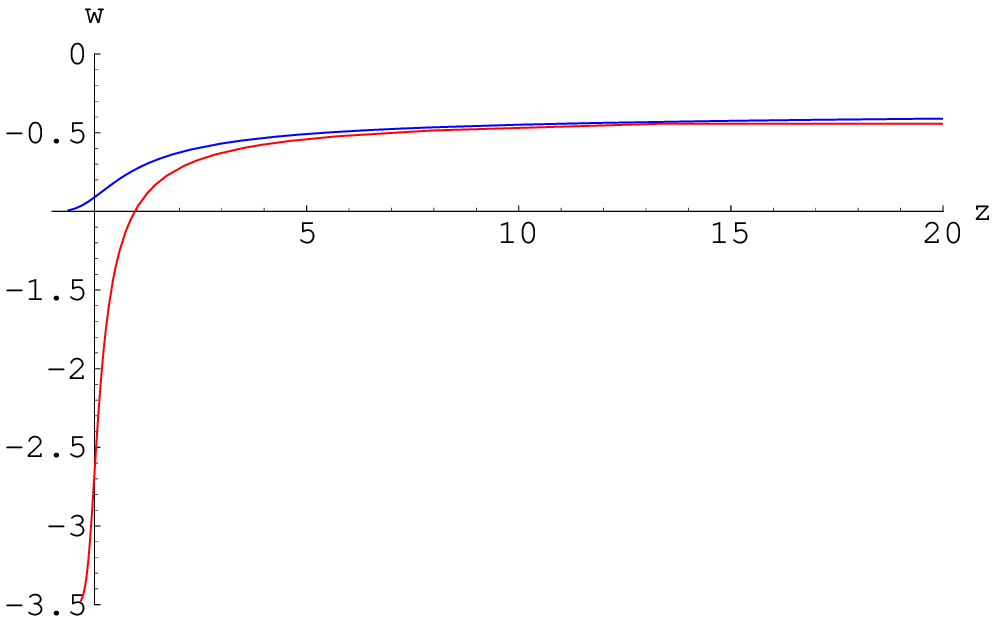}}
        \nobreak\bigskip
    {\raggedright\it \vbox{
{\bf Figure 3.} {\it The evolution of $w = w_\d (z)$,
here the blue line corresponds to
$\Omega^0_\d = 0.75$, $d=1$ and the red one corresponds to
$\Omega_\d^0=0.54$, $d=0.21$.}
 }}}}
    \bigskip}

\bigskip
{\vbox{{\epsfxsize=8cm
        \nobreak
    \centerline{\epsfbox{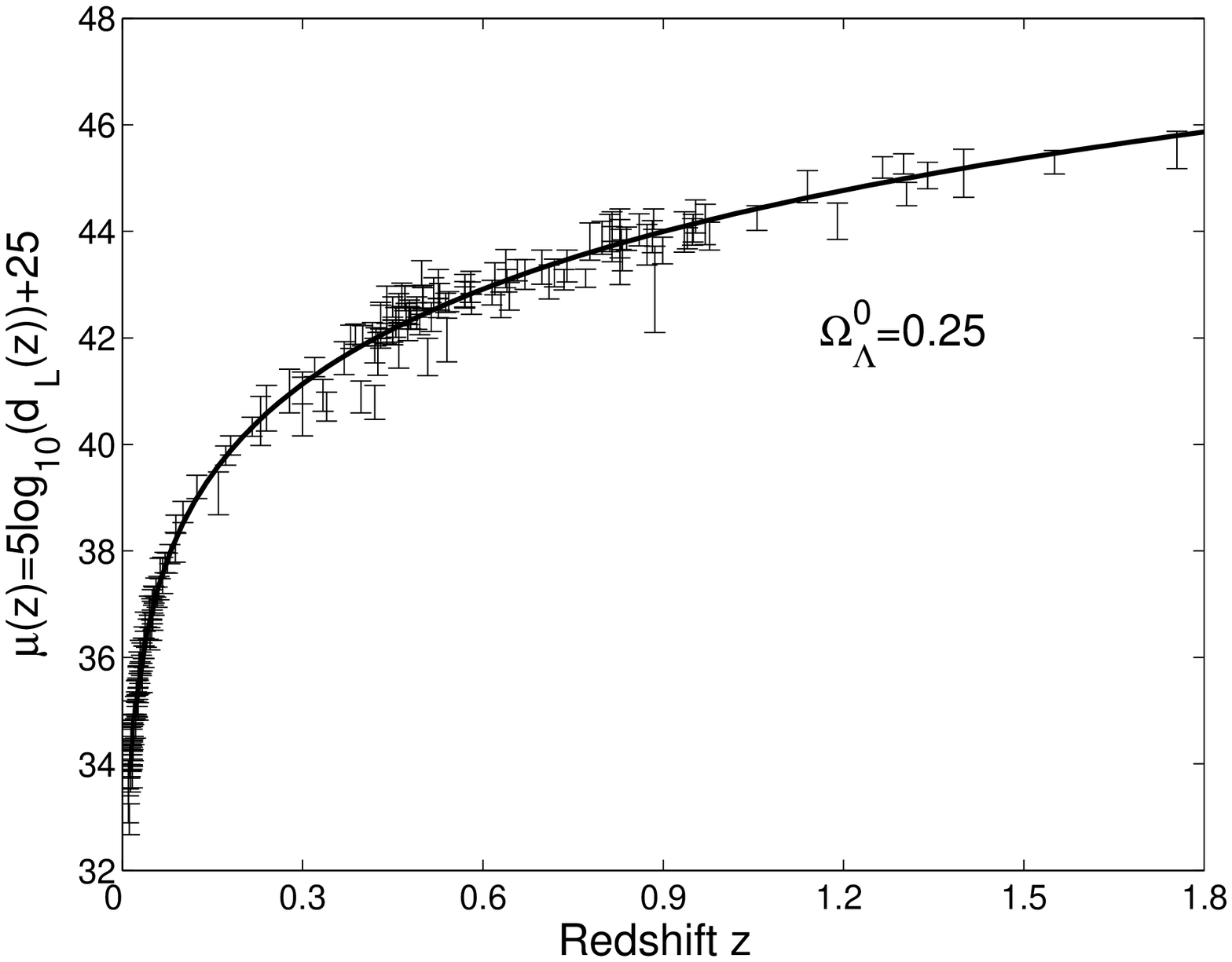}}
        \nobreak\bigskip
    {\raggedright\it \vbox{
{\bf Figure 4.} {\it The relation between the extinction-corrected
distance moduli and redshift for the best fit model, here $d=1$. }
 }}}}
    \bigskip}

In \spna, Riess \etal\ found that $w^0_\d<-0.76$ at the 95\%
confidence level by using SN Ia data, our result $w^0_\d=-0.91$
with $d=1$ and $w^0_\d=-2.67$ with $d=0.21$ are consistent with
that. Recently, Tegmark \etal\ found that $\Omega^0_m\approx
0.30\pm 0.04$ by using the Wilkinson Microwave Anisotropy Probe
(WMAP) data in combination with the Sloan Digital Sky Survey
(SDSS) data \tegmark. If we use this prior, then we find the best
fit parameters are $\Omega_m^0=0.32\pm 0.06$ and
$d=0.64^{+0.36}_{-0.24}$ with $\chi^2=175.93$. Then the transition
redshift is $z_T=0.53$. The plot in Fig. 3 is consistent with the
model independent analysis over the evolution of $\Omega_\d$ by
using the WMAP and SN Ia data in \gong. More recent model
independent analysis favor a phantom like dark energy model and
lower transition redshift $z_T\sim 0.3$ or $z_T\sim 0.4$ \indep.
The two parameter representation of dark energy models also favor
a higher value for $\Omega_m^0$. Our best fit result is consistent
with those analysis.

In conclusion, the holographic dark energy model is consistent
with current observations and the more precise cosmological
observations will be taken to be the decided constraints on this
model. The model is also a better fit to observations than the
$\d$CDM model.

\bigskip

Acknowledgments.

We would like to thank the anonymous referee for fruitful
comments. We thank Miao Li for his comment on the manuscript.
Qing-Guo Huang would like to thank the Interdisciplinary Center
for Theoretical Study at University of Science and Technology of
China for hospitality during the course of this work. Y. Gong is
supported by CQUPT under grants A2003-54 and A2004-05.

\listrefs
\end

)
\input harvmac
\input epsf

\def\d{\Lambda}

\def\etal{ et al.}



\Title{}{\vbox{\centerline{Supernova Constraints on a holographic
dark energy model}}}

\centerline{Qing-Guo Huang$^{1,2}$ and Yungui Gong$^{3}$ }
\medskip
\centerline{$^{1}$\it Institute of Theoretical Physics}
\centerline{\it Academia Sinica, P.O. Box 2735} \centerline{\it
Beijing 100080}
\medskip
\centerline{$^{2}$\it Interdisciplinary Center for Theoretical
Study,} \centerline{\it University of Science and Technology of
China,} \centerline{\it Hefei, Anhui 230026, P. R. China}
\medskip
\centerline{$^{3}$\it Center for Relativity and Astrophysics and
College of Electronic Engineering, } \centerline{\it Chongqing
University of Posts and Telecommunications, } \centerline{\it
Chongqing 400065, China}
\medskip
\centerline{\tt huangqg@itp.ac.cn} \centerline{\tt
gongyg@cqupt.edu.cn}
\bigskip

In this paper, we use the type Ia supernova data to constrain the
model of holographic dark energy. For $d=1$, the best fit result
is $\Omega^0_m = 0.25$, the equation of the state of the
holographic dark energy $w^0_\d = -0.91$ and the transition
between the decelerating expansion and the accelerating expansion
happened when the cosmological red-shift was $z_T=0.72$. If we set
$d$ as a free parameter, the best fit results are $d=0.21$,
$\Omega_m^0=0.46$, $w_\d^0=-2.67$, which sounds like a phantom
today, and the transition redshift is $z_T=0.28$.

\Date{March, 2004}

\nref\agr{A. G. Riess et al., Astron. J. 116 (1998) 1009; S.
Perlmutter et al., APJ 517 (1999) 565. }

\nref\tonbar{J. L. Tonry et al., APJ 594 (2003) 1; R. A. Knop et
al., astro-ph/0309368; B. J. Barris et al., APJ 60 (2004) 571.}

\nref\spna{A. G. Riess et al, astro-ph/0402512. }

\nref\cmb{P. de Bernardis et al., Nature 404 (2000) 955; S. Hanany
et al., APJ 545 (2000) L5; D. N Spergel et al., APJ Supp. 148
(2003) 175.}

\nref\dark{C. Wetterich, Nucl. Phys. B 302 (1988) 668; B. Ratra
and P. J. E. Peebles, Phys. Rev. D 37 (1988) 3406; R. R. Caldwell,
R. Dave and P. J. Steinhardt, Phys. Rev. Lett. 80 (1998) 1582; I.
Zlatev, L. Wang and P. J. Steinhardt, Phys. Rev. Lett. 82 (1999)
896.}

\nref\review{ V. Sahni and A. A. Starobinsky, Int. J. Mod. Phys. D
9 (2000) 373; T. Padmanabhan, Phys. Rep. 380 (2003) 235.}

\nref\thooft{ G. 't Hooft, Salamfestschrift: a collection of
talks, Eds. A. Ali, J. Ellis and S. Randjbar-Daemi, World
Scientific, 1993.}

\nref\susskind{ L. Susskind, J. Math. Phys. 36 (1994) 6377.}

\nref\ckn{A. Cohen, D. Kaplan and A. Nelson, Phys. Rev. Lett. 82
(1999) 4971.}

\nref\hsu{S. D. H. Hsu, hep-th/0403052. }

\nref\horvat{R. Horvat, astro-ph/0404204.}

\nref\ph{W. Fischler and L. Susskind, hep-th/9806039; N. Kaloper
and A. Linde, Phys. Rev. D 60 (1999) 103509; R. Bousso, Rev. Mod.
Phys. 74 (2002) 825. }

\nref\ml{Miao Li, hep-th/0403127. }

\nref\tbf{T. Banks and W. Fischler, astro-ph/0307459.}

\nref\tegmark{M. Tegmark et al., Phys. Rev. D 69 (2004) 103501.}

\nref\gong{Y. Gong, astro-ph/0401207.}

\nref\indep{R. A. Daly and S. G. Djorgovski, astro-ph/0403664; U.
Alam, V. Sahni, T.D. Saini and A.A. Starobinsky, astro-ph/0311364;
U. Alam, V. Sahni and A. A. Starobinsky, astro-ph/0403687; H. K.
Jassal, J. S. Bagla and T. Padmanabhan, astro-ph/0404378; Y. Gong,
astro-ph/0405446.}

The type Ia supernova (SN Ia) observations \refs{\agr, \tonbar,
\spna} provide evidence that the expansion of our universe at the
present time appears to be accelerating which is due to dark
energy with negative pressure. The kinematic interpretation of the
relationship between SN Ia luminosity distance and red-shift is
most consistent with two distinct epoches of expansion: a recent
accelerated expansion and a previous decelerated expansion with a
transition around $z_{\rm T}\sim 0.4$ between them \spna. This is
a generic requirement of a mixed dark matter and dark energy
universe. The cosmic background microwave (CMB) observations \cmb\
hint a spatially flat universe. However, the unusual small value
of the cosmological constant, which is extremely smaller than the
estimate from the effective local quantum field theory, is perhaps
one of the biggest puzzles and deepest mysteries in modern
physics. Alternatively, many dynamical dark energy models with
cosmological constant like behavior were proposed in the
literature \dark\ and \review. For a review, see, for example
\review\ and references therein.

't Hooft \thooft\ and Susskind \susskind\ showed that the
effective local quantum field theories greatly over-count degrees
of freedom because the entropy scales extensively for an effective
quantum field theory in a box of size $L$ with UV cutoff $\d$. In
order to solve the problem, A. Cohen et al. \ckn\ proposed a
relationship between UV and IR cut-offs corresponding to the
assumption that the effective field theory describes all states of
the system excluding those for which have already collapsed to a
black hole. If the sum of the zero-point energies of all normal
modes of the fields is $\rho_\d$, we must have $L^3 \rho_\d \leq L
M_p^2$ or $\rho_\d \leq M_p^2 L^{-2}$, this means that the maximum
entropy is in the order of $S^{3/4}_{BH}$. The magnitude of the
holographic energy proposed by Cohen et al. may be the same as
that from cosmological observations. But Hsu recently pointed out
that the equation of state is not correct for describing the
accelerating expansion of our Universe in \hsu. In other words,
the original holographic energy couldn't give an accelerating
universe. The idea was later generalized to make the gravitational
constant varying with time in \horvat.

The origin of the Bekenstein-Hawking constraint on the entropy of
a black hole is the existence of the event horizon, which serves
as a natural boundary for all processes inside a black hole.
However, there is no event horizon in a non-inflationary universe
and we should replace it with the particle horizon, which has been
discussed in \ph. But there is an event horizon in the Universe
with accelerating expansion and it is a natural choice that the
event horizon acts as the boundary of the Universe. Very recently,
Li suggested that we should use the proper future event horizon of
our Universe to cut-off the large scale and bring about an
accelerating expansion of our Universe in \ml. On the other hand,
Banks and Fischler have pointed out that the the number of the
e-floldings during inflation is bounded, which is due to the bound
on the entropy, if we take the event horizon as the boundary of
our Universe and the present acceleration of the Universe is due
to an asymptotically de Sitter universe with small cosmological
constant in \tbf.

In this paper, we use the new SN Ia data compiled by Riess \etal\
to constrain the holographic dark energy model proposed by Li.
Firstly we take a short trip on the holographic energy model
proposed in \ml. According to \ml, the energy density of the
holographic dark energy is \eqn\erd{\rho_\d = 3 d^2 M^2_p
R^{-2}_h, } here we keep $d$ as a free parameter (the author of
\ml\ favored $d=1$) and $R_h$ is the proper size of the future
event horizon, \eqn\eh{R_h(t) = a(t) \int^{\infty}_{t} {dt' \over
a(t')} = a \int^{\infty}_{a} {da' \over H' a'^2}. } For a
spatially flat, isotropic and homogeneous universe with an
ordinary matter and dark energy, the Friedmann equation is
\eqn\fde{ \Omega_\d + \Omega_m = 1, \quad \Omega_m = {\rho_m \over
\rho_{cr}} \quad \hbox{and} \quad \Omega_\d={\rho_\d \over
\rho_{cr}}, } where $\rho_m$ ($\rho_\d$) is the energy density of
matter (dark energy) and the critical density $\rho_{cr} = 3 M^2_p
H^2$. Using Friedmann equation \fde\ and $\rho_m = \rho^0_m
a^{-3}=3 M^2_p H^2_0 \Omega_m^0 a^{-3}$, where we set $a_0 =1$ and
$a = (1 + z)^{-1}$, we have $\Omega_m=1-\Omega_\d=(H_0/H)^2
\Omega_m^0 a^{-3}$,
 \eqn\ah{{1 \over a H}
= a ^{1/2} {1 \over \sqrt{\Omega^0_m} H_0} ( 1 - \Omega_\d
)^{1/2}, } and \eqn\rd{\rho_\d = \Omega_\d \rho_{cr} = {\Omega_\d
\over 1 - \Omega_\d} \Omega_m \rho_{cr} = {\Omega_\d \over 1 -
\Omega_\d} \rho_m = {\Omega_\d \over 1 - \Omega_\d} \rho^0_m
a^{-3}. } Combining equations \erd\ and \rd, we find
\eqn\rhm{R_h(t) = a^{3/2} {d \over \sqrt{\Omega^0_m} H_0} \left({1
- \Omega_\d \over \Omega_\d} \right)^{1/2}. } Substituting
equations \ah\ and \rhm\ into \eh, \eqn\inq{\int^{\infty}_{x} dx'
e^{x'/2} (1 - \Omega_\d)^{1/2} = d e^{x/2} \left({1 - \Omega_\d
\over \Omega_\d} \right)^{1/2}, } where $x = \ln a$. Taking
derivative with respect to $x$ in both sides of equation \inq, we
get \eqn\dnq{\Omega'_\d = \Omega_\d (1 - \Omega_\d) ( 1 + {2 \over
d} \sqrt{\Omega_\d} ), } where the prime denotes the derivative
with respect to $x$. We can get the analytic solution of equation
\dnq\ as, \eqn\mz{\ln \Omega_\d - {d \over 2+d} \ln (1 -
\sqrt{\Omega_\d}) + {d \over 2-d} \ln (1 +\sqrt{\Omega_\d}) - {8
\over 4-d^2} \ln(d + 2 \sqrt{\Omega_\d}) = - \ln(1 + z) + y_0, }
where $y_0$ can be determined by the value of $\Omega^0_\d$
through equation \mz.

Because of the conservation of the energy-momentum tensor, the
evolution of the energy density of dark energy is governed by
\eqn\eed{{d \over da} (a^3 \rho_\d) = -3 a^2 p_\d. } Thus we
obtain \eqn\pd{p_\d = - {1 \over 3} {d \rho_\d \over d \ln a} -
\rho_\d. } Using equations \rd\ and \dnq, after a lengthy but
straightforward calculation, we find the pressure of dark energy
can be expressed as \eqn\pd{p_\d = - {1 \over 3} (1 + {2 \over d}
\sqrt{\Omega_\d}) \rho_\d, } and the equation of state of dark
energy is \eqn\md{w_\d = {p_\d \over \rho_\d} = - {1 \over 3} (1 +
{2\over d} \sqrt{\Omega_\d}), } and  \eqn\rmd{{d w_\d \over d z} =
{1 \over 3d} \sqrt{\Omega_\d} (1 - \Omega_\d) ( 1 + {2\over d}
\sqrt{\Omega_\d} ) {1 \over 1 + z}. } Since $0 \leq \Omega_\d \leq
1$, we find the equation of state of dark energy $-(1+2/d)/3 \leq
w_\d \leq -{1\over 3}$ and the evolution of $w_\d$ is slow. If we
use $\Omega^0_\d = 0.73$ and $d=1$, we obtain $w^0_\d = - 0.90$
and $dw^0_\d/dz=0.21$. In the future, our Universe will be
dominated by dark energy with $w_\d = -(1+2/d)/3$, this result is
the same as Eq. (9) in \ml. The expansion of our Universe will be
accelerating forever.

In the past, the expansion of our Universe experienced
deceleration due to domination by radiation or matter. With the
evolution, our Universe will be dominated by the dark energy and
the expansion of our Universe starts to be accelerating. This
transition happened when \eqn\acc{{\ddot a \over a} = - {1 \over 6
M^2_p} (\rho_\d + 3 p_\d + \rho_m) = 0.}  Using equations \rd\ and
\pd, we obtain \eqn\tran{\Omega^T_\d+{2\over d}\Omega^T_\d
\sqrt{\Omega^T_\d}=1. } If $d=1$, solving this equation, we find
that the turning point is corresponding to $\Omega^T_\d=0.4320$.
For $d \rightarrow \infty$, $\Omega_\d^T \simeq 1$. If
$\Omega_\d^0$ is finite, $y_0$ in equation \mz\ will be finite.
Using equation \mz\ again, we obtain the red-shift of turning
point must be $z_T \simeq -1$. This result can be understood
easily. Since $d \rightarrow \infty$ and $\Omega_\d$ must be
finite, $w_\d \rightarrow -1/3$. But the energy density of matter
will be red-sifted faster than the holographic dark energy. So in
the far future our Universe will be dominated by the dark energy
and its expansion will be accelerating because of $w_\d < -1/3$.
On the other hand, for $d \rightarrow 0$, equation \tran\ tells us
\eqn\dtran{\Omega_\d^T \simeq (d/2)^{2/3}.} Also assuming
$\Omega_\d^0$ is finite, applying equation \mz, we have $y_0 = -2
\ln 2$. Substituting equation \dtran\ into \mz, we obtain the
red-shift of turning point is $z_T \simeq 0$. In this case, $w_\d
\rightarrow - \infty$ for finite $\Omega_\d$ and the energy
density of the holographic dark energy will increase very fast.
Therefore our Universe was dominated by this dark energy and its
expansion started to be accelerating very recently, if
$\Omega_\d^0$ is finite. We show the relation between the
red-shift of the turning point and the value of the parameter $d$
for some finite $\Omega_\d^0$ in Fig. 1. This figure is consistent
with our previous analysis.
\bigskip
{\vbox{{\epsfxsize=8cm
        \nobreak
    \centerline{\epsfbox{tzd.eps}}
        \nobreak\bigskip
    {\raggedright\it \vbox{
{\bf Figure 1.} {\it $z = z_T$ is the cosmological red-shift
corresponding to the turning point between decelerating expansion
and accelerating expansion, here the blue line corresponds to
$\Omega^0_\d = 0.75$ and the red one corresponds to
$\Omega_\d^0=0.54$.}
 }}}}
    \bigskip}

The luminosity distance $d_{\rm L}$ expected in a spatially flat
Friedmann - Robertson - Walker (FRW) cosmology with mass density
$\Omega_m$ and the holographic dark energy density $\Omega_\d$ is
\eqn\lumin{\eqalign{ d_{\rm L}(z)=&c(1+z)\int^{t_0}_t {d t'\over
a(t')}=c(1+z)[(1+z)R_h(t)-R_h(t_0)] \cr
=&c\,d\,H^{-1}_0\left[\sqrt{{1-\Omega_\Lambda\over \Omega_\Lambda
\Omega_m^0}}(1+z)^{1/2}-(1-\Omega^0_m)^{-1/2}(1+z)\right] .}} In
the above derivation, we used Eqs. \eh and \rhm. The parameter
$\Omega^0_m$ of the model and the nuisance parameter $H_0$ are
determined by minimizing \eqn\lrmin{ \chi^2=\sum_i{[\mu_{\rm
obs}(z_i)-\mu(z_i)]^2\over \sigma^2_i},} where the
extinction-corrected distance moduli $\mu(z)=5\log_{10}(d_{\rm
L}(z)/{\rm Mpc})+25$ and $\sigma_i$ is the total uncertainty in
the observation. The nuisance parameter $H_0$ is marginalized. If
$d=1$, the best fit to the 157 gold SN sample in \spna\ is
$\Omega_m^0=0.25^{+0.04}_{-0.03}$ with $\chi^2=176.7$ or
$\chi^2/{\rm dof}=1.133$, and the best fit to the whole 186 gold
and silver SN sample is $\Omega_m^0=0.25\pm{0.03}$ with
$\chi^2=232.8$ or $\chi^2/{\rm dof}=1.258$. By using the best fit
$\Omega_m^0$, we find that $w^0_\d=-0.91 \pm 0.01$ and the
red-shift corresponding to transition is
$z_T=0.72^{+0.11}_{-0.13}$. For comparison, the best fit to the
gold SN sample for the $\d$-model is $\Omega_m^0=0.31\pm 0.04$
with $\chi^2=177.1 $ or $\chi^2/{\rm dof}=1.135$. So the
transition redshift for $\d$-model is $z_T=0.65^{+0.11}_{-0.10}$.
If we set $d$ as a free parameter, we find the best fit to the
gold SN sample is $\Omega_m^0=0.46^{+0.08}_{-0.13}$ and
$d=0.21^{+0.45}_{-0.14}$ with $\chi^2=173.45$ or $\chi^2/{\rm
dof}=1.119$. In this case the red-shift corresponding to
transition is $z_T=0.28^{+0.23}_{-0.13}$. The best fit contour for
$\Omega_m^0$ and $d$ is plotted in Fig. 2. The best fit to the
gold and silver SN sample is $\Omega_m^0=0.46^{+0.14}_{-0.11}$ and
$d=0.20^{+0.28}_{-0.10}$ with $\chi^2=226.4$ or $\chi^2/{\rm
dof}=1.230$. And the red-shift corresponding to transition is
$z_T=0.27^{+0.19}_{-0.14}$.

\bigskip
{\vbox{{\epsfxsize=8cm
        \nobreak
    \centerline{\epsfbox{holocf.eps}}
        \nobreak\bigskip
    {\raggedright\it \vbox{
{\bf Figure 2.} {\it The best fit contour for $\Omega_m^0$ and $d$
to the gold sample SNe.}
 }}}}
    \bigskip}

Using the fit parameters $\Omega^0_\d = 0.75$ and $d=1$, we get
$y_0 = -1.67$. With the best fit parameters $\Omega^0_\d = 0.54$
and $d=0.21$, we get $y_0 = -1.47$. Combining equations \mz, \md\
and \lumin, we show the equation of state of the dark energy
$w_\d$ in Fig. 3 and the extinction-corrected distance moduli in
Fig. 4.
\bigskip
{\vbox{{\epsfxsize=8cm
        \nobreak
    \centerline{\epsfbox{mdz.eps}}
        \nobreak\bigskip
    {\raggedright\it \vbox{
{\bf Figure 3.} {\it The evolution of $w = w_\d (z)$,
here the blue line corresponds to
$\Omega^0_\d = 0.75$, $d=1$ and the red one corresponds to
$\Omega_\d^0=0.54$, $d=0.21$.}
 }}}}
    \bigskip}

\bigskip
{\vbox{{\epsfxsize=8cm
        \nobreak
    \centerline{\epsfbox{ddz.eps}}
        \nobreak\bigskip
    {\raggedright\it \vbox{
{\bf Figure 4.} {\it The relation between the extinction-corrected
distance moduli and redshift for the best fit model, here $d=1$. }
 }}}}
    \bigskip}

In \spna, Riess \etal\ found that $w^0_\d<-0.76$ at the 95\%
confidence level by using SN Ia data, our result $w^0_\d=-0.91$
with $d=1$ and $w^0_\d=-2.67$ with $d=0.21$ are consistent with
that. Recently, Tegmark \etal\ found that $\Omega^0_m\approx
0.30\pm 0.04$ by using the Wilkinson Microwave Anisotropy Probe
(WMAP) data in combination with the Sloan Digital Sky Survey
(SDSS) data \tegmark. If we use this prior, then we find the best
fit parameters are $\Omega_m^0=0.32\pm 0.06$ and
$d=0.64^{+0.36}_{-0.24}$ with $\chi^2=175.93$. Then the transition
redshift is $z_T=0.53$. The plot in Fig. 3 is consistent with the
model independent analysis over the evolution of $\Omega_\d$ by
using the WMAP and SN Ia data in \gong. More recent model
independent analysis favor a phantom like dark energy model and
lower transition redshift $z_T\sim 0.3$ or $z_T\sim 0.4$ \indep.
The two parameter representation of dark energy models also favor
a higher value for $\Omega_m^0$. Our best fit result is consistent
with those analysis.

In conclusion, the holographic dark energy model is consistent
with current observations and the more precise cosmological
observations will be taken to be the decided constraints on this
model. The model is also a better fit to observations than the
$\d$CDM model.

\bigskip

Acknowledgments.

We would like to thank the anonymous referee for fruitful
comments. We thank Miao Li for his comment on the manuscript.
Qing-Guo Huang would like to thank the Interdisciplinary Center
for Theoretical Study at University of Science and Technology of
China for hospitality during the course of this work. Y. Gong is
supported by CQUPT under grants A2003-54 and A2004-05.

\listrefs
\end